\documentclass[aps,preprint,amsmath,amssymb,prb,superscriptaddress,unsortedaddress]{revtex4}
\usepackage{epsf}
\usepackage{graphicx}
\usepackage{sidecap}

\begin{document}

\title{Spin correlated electronic state on the surface of a spin-orbit Mott system}

\author{Chang~Liu}
\affiliation {Joseph Henry Laboratory and Department of Physics, Princeton University, Princeton, New Jersey 08544, USA}

\author{Su-Yang~Xu}
\affiliation {Joseph Henry Laboratory and Department of Physics, Princeton University, Princeton, New Jersey 08544, USA}

\author{Nasser~Alidoust}
\affiliation {Joseph Henry Laboratory and Department of Physics, Princeton University, Princeton, New Jersey 08544, USA}

\author{Tay-Rong~Chang}
\affiliation {Department of Physics, National Tsing Hua University, Hsinchu 30013, Taiwan}

\author{Hsin~Lin}
\affiliation {Department of Physics, Northeastern University, Boston, Massachusetts 02115, USA}

\author{Chetan~Dhital}
\affiliation{Department of Physics, Boston College, Chestnut Hill, Massachusetts 02467, USA}

\author{Sovit~Khadka}
\affiliation{Department of Physics, Boston College, Chestnut Hill, Massachusetts 02467, USA}

\author{Madhab~Neupane}
\affiliation {Joseph Henry Laboratory and Department of Physics, Princeton University, Princeton, New Jersey 08544, USA}

\author{Ilya~Belopolski}
\affiliation {Joseph Henry Laboratory and Department of Physics, Princeton University, Princeton, New Jersey 08544, USA}

\author{Gabriel~Landolt}
\affiliation {Swiss Light Source, Paul Scherrer Institut, CH-5232, Villigen, Switzerland}

\author{Horng-Tay~Jeng}
\affiliation {Department of Physics, National Tsing Hua University, Hsinchu 30013, Taiwan}

\author{Robert~S.~Markiewicz}
\affiliation {Department of Physics, Northeastern University, Boston, Massachusetts 02115, USA}

\author{J.~Hugo~Dil}
\affiliation {Swiss Light Source, Paul Scherrer Institut, CH-5232, Villigen, Switzerland}
\affiliation {Physik-Institut, Universit\"{a}t Z\"{u}rich-Irchel, 8057 Z\"{u}rich, Switzerland}

\author{Arun~Bansil}
\affiliation {Department of Physics, Northeastern University, Boston, Massachusetts 02115, USA}

\author{Stephen~D.~Wilson}
\affiliation {Department of Physics, Boston College, Chestnut Hill, Massachusetts 02467, USA}

\author{M.~Zahid~Hasan}
\affiliation {Joseph Henry Laboratory and Department of Physics, Princeton University, Princeton, New Jersey 08544, USA}

\date{\today}

\clearpage

\begin{abstract}

Novel phases of two dimensional electron systems resulting from new surface or interface modified electronic structures have generated significant interest in material science. We utilize photoemission spectroscopy to show that the near-surface electronic structure of a bulk insulating iridate Sr$_3$Ir$_2$O$_7$ lying near metal-Mott insulator transition exhibit weak metallicity signified by finite electronic spectral weight at the Fermi level. The surface electrons exhibit a unique spin structure resulting from an interplay of spin-orbit, Coulomb interaction and surface quantum magnetism, distinct from a topological insulator state. Our results suggest the experimental realization of a novel quasi two dimensional interacting electron surface ground state, opening the door for exotic quantum entanglement and transport phenomena in iridate-based oxide devices.

\end{abstract}


\maketitle

Strongly correlated electronic behavior can be modified near surfaces and interfaces of transition metal oxides, leading to novel quantum phenomena \cite{Hwang, Syro_Nature, Thiel_Science}. Surfaces are known to be dramatically modified in topological band insulators and in spin-orbit coupled Rashba semiconductors \cite{Moore_Nature, Suyang_NPhys, BiTeI_1}. These unusual surface effects not only reflect novel physics but also hold potential for future devices where such effects are amplified by nano-structuring, which leads to the enhancement of surface-to-bulk ratio \cite{Paglione}. Recently, attention has focused on materials in which Mott physics and strong spin-orbit interaction may coexist in the bulk. Iridium oxides (iridates) have been identified to be one of such promising classes of materials \cite{Cao_327, BJKim_Rotenberg_PRL, Moon_PRL, Dessau, Baumberger, Wojek, Jackeli, Shitade, Sakakibara_Nature, Damascelli}. So far, research on the iridates has largely focused on their bulk properties. Theoretical models suggest the possibility of realizing exotic phenomena in the iridates, such as the arced semimetal \cite{Wan_Weyl}, topological insulator \cite{YBKim_TI}, and high temperature superconductivity \cite{Senthil}, none of which has yet been found experimentally. Here we report a different route to look for exotic surface phenomena in iridates and identify the surface electronic and spin ground state. Surface phenomena are often expected to be enhanced in correlated systems near the bulk metal-Mott insulator transition at which Coulumb interaction, spin-orbit coupling and the often-frustrated magnetic moments compete in determining the ground state. Near such metal-insulator bulk criticality, surface modification is likely to occur since the narrow-gap bulk states are sensitive to changes and relaxations of surrounding crystal potential near the surface. The iridate we focus here, Sr$_3$Ir$_2$O$_7$, belongs to the Ruddlesden-Popper series whose bulk electronic structure lies in between a Mott insulator ($n = 1$) and a correlated metal ($n = \infty$) \cite{Moon_PRL, Cao_327}, in the vicinity of a bulk criticality. Evidently, it is critically important to identify the surface ground state of such exotic iridates if they differ from the bulk.

We report a systematic study of spin integrated and spin resolved angle resolved photoemission spectroscopy (ARPES) to critically (and thoroughly) investigate the near-surface electronic structure of Sr$_3$Ir$_2$O$_7$. While the bulk exhibits strongly insulating transport properties in this compound, our photoemission results reveal finite spectral weight at the Fermi level and a gap-like suppression for quasiparticles within 30-40 meV of the Fermi level. In addition, the low energy electrons exhibit strong left-right imbalanced modulation related to surface spin polarization, as well as a unique spin fine structure which reveals an in-plane Rashba-like spin polarization induced by onsite Coulomb interaction on the surface. Such a rich character of the surface ground state is not expected within the calculated and predicted bulk electronic structure. We show that this material exhibits a critical interplay of spin-orbit coupling, antiferromagnetism, and surface termination. Our results point toward the experimental realization of a new type of correlated surface electron system on the boundary of a bulk material lying in the critical ($U \sim W$) regime.

Fig. \ref{Fig1} shows the ARPES band dispersion data within the $k_x$-$k_y$ plane [(001) plane]. At low temperatures, Sr$_3$Ir$_2$O$_7$ changes from a paramagnetic phase to an antiferromagnetic (AF) phase, where the Ir moments point along the $c$-axis \cite{JWKim} with an in-plane commensurate N\'{e}el vector \cite{Boseggia}. In Fig. \ref{Fig1}\textbf{a} we plot the raw and differentiated in-plane resistivity as a function of temperature (adapted from Ref. \onlinecite{Wilson_new}). The ordering temperature $T_{\mathrm{AF}} \sim 280$ K is clearly shown as a sharp peak in the $\frac {\partial \mathrm{log} \rho}{\partial (1/T)}$ vs. $T$ curve \cite{Wilson_new}; a drastic upturn of the $\rho$ vs. $T$ curve at low temperatures confirms its bulk insulating transport. Figs. \ref{Fig1}\textbf{b} and \textbf{c} summarize the in-plane electronic structure obtained from ARPES (Figs. \ref{Fig1}\textbf{d}-\textbf{f}). In surprising contrast with the transport data (Fig. \ref{Fig1}\textbf{a}), two bands ($\alpha$, blue; $\beta$, green) evolve to a close vicinity of the Fermi level. The shape and dispersive pattern of these low-lying bands are rather complicated. Fig. \ref{Fig1}\textbf{d} shows a typical ARPES Fermi mapping obtained with 35 eV photons. Finite spectral weight is present at the Fermi level ($E_F$), indicative of a nearly conductive ground state. This finite intensity at $E_F$ can be due to possible band bending effect close to the sample surface. Such a band bending effect - strong enough so that $E_F$ descends to near the bottom of the Mott gap at a depth greater than the electron escape depth - will give rise to an ARPES signal dominated by the more conductive surface layer. The fact that this conducting channel is not detected by conventional transport measurements may be due to a small surface/bulk volume ratio and difference in mobility between the bulk- and surface-originated charge carriers. Nonetheless, this is a surface modified effect. At 0.15 eV binding energy (bottom panel of Fig. \ref{Fig1}\textbf{e}), the $\alpha$-band decomposes into segments; the $\beta$-band shrinks in size despite maintaining intact, signifying its electronlike nature. It is important to point out here that the $\beta$ band is not observed in previous ARPES works on this material \cite{Dessau, Baumberger, Wojek}, possibly because of different sample quality and our choice of measuring the second Brillouin zone instead of the first, where no sign of the $\beta$ band is seen. In Fig. \ref{Fig1}\textbf{f} we show three $k$-$E$ maps along directions shown in Fig. \ref{Fig1}\textbf{b}. It should be noted from Cuts 1 and 2 that, despite the finite intensity at $E_F$, the quasiparticle structure is gapped at $E_F$ (Fig. \ref{Fig1}\textbf{g}). The low energy quasiparticles experience a gap-like spectral weight suppression (SWS), signified by a gradual decrease of ARPES intensity as the bands approaching the Fermi level, similar to other correlated oxides \cite{Chuang_Science}. The second derivative analysis of Cut 1 (Ref. \onlinecite{Zhang_deri}) (bottom panels) reveals that, along the $\Gamma$-$X$ direction, the $\alpha$ band bends horizontally to form a van Hove-like flat portion (Supplementary Information), while the $\beta$ band loses its intensity. Fig. \ref{Fig1}\textbf{g} shows the existence of the SWS at the $X$ point where the $\alpha$-band evolves closest to $E_F$ (i.e., top of the flat portion). The exact location of the Fermi level is obtained by fitting the polycrystalline gold data with the Fermi distribution function; the EDC at the $X$ point is then symmetrized with respect to $E_F$. The two peak and valley lineshape of the symmetrized EDC proves the presence of the SWS, which is about 38 meV in size. This value is smaller than the full insulating gap value obtained from optical measurements ($\sim 250$ meV, Ref. \onlinecite{Moon_PRL}), indicative of a different $E_F$ at the crystal surface resulting from effective band bending. We choose to study the least resistive samples here since (1) these are the only samples who show the presence of the $\beta$ band, which is the main focus of our spin resolved ARPES study, (2) these samples do not show charging effect, and (3) the 38 meV gap observed in these samples does not change with temperature (Supplementary Information). Note that concurrent STS results \cite{Yoshi} on the same samples used in our studies consistently supports our observation that the surface Fermi level of Sr$_3$Ir$_2$O$_7$ always lies close to the top of the valence bands. To sum up our results in Fig. \ref{Fig1}, we observe a gap-like suppression of spectral weight for the spin-orbit correlated electrons at the surface of Sr$_3$Ir$_2$O$_7$ within a narrow energy window of $E_F$, as well as an electronlike Fermi contour surrounding $\Gamma$ (the $\beta$ band) which has not been observed in previous studies \cite{Dessau, Baumberger, Wojek}.

In Fig. \ref{Fig2} we present the spin resolved (SR) ARPES measurements of Sr$_3$Ir$_2$O$_7$. A spin-integrated ARPES $k$-$E$ map along the $X$-$\Gamma$-$X$ direction (see also Fig. \ref{Fig1}\textbf{f}) is shown in Fig. \ref{Fig2}\textbf{a} for clarification of momentum space positions and band notations. The important observations of this data set are (1) the strong left-right imbalance of the SR-ARPES signal, and (2) the Rashba-like spin fine structure of the $\beta$ band. We summarize our results in Fig. \ref{Fig2}\textbf{b}, where the in-plane spin polarization angles at different $k$-points are presented \cite{Hugo} (numbers in Fig. \ref{Fig2}\textbf{b}). It is seen from Fig. \ref{Fig2}\textbf{b} that both the $\alpha$ and the $\beta$ band are spin polarized. The integrated spin for the $\alpha$ band near the left and the right $X$ momenta point to opposite directions. The $\beta$ band consists of two close-by rings with Rashba-like in-plane spin helicity. In Figs. \ref{Fig2}\textbf{c}-\textbf{e} we show the spin polarization analysis for a MDC cut along the $X$-$\Gamma$-$X$ direction at $\sim50$ meV binding energy, within the same energy window as the SWS observed in Fig. \ref{Fig1} (see Supplementary Information for more details). The spin direction under study is along $[1,-1,0]$, which is tangential to the $\beta$ contour. First, multiple peaks are present in the total intensity curve (blue circles). The spin-up and spin-down components show strong antisymmetry with respect to the zone center $\Gamma$. Since this behavior contradicts the Kramers' theorem, one possible reason for its occurrence is the presence of surface antiferromagnetism (AFM) which explicitly breaks time reversal symmetry, although we cannot rule out other many-body effects which give rise to a collective net spin polarization at the surface. These effects introduce a spin polarized background (blue curve in the inset of Fig. \ref{Fig2}\textbf{d}); spin polarized signals from individual bands sit on top of this background. Second, the spin fine structure of the $\beta$ band is observed in the raw polarization curve (Fig. \ref{Fig2}\textbf{d}). More specifically, the $\beta$ band consists of two close-by rings with Rashba-like opposite in-plane spin helicity. At around $k = 0.5$ $\mathrm{\AA}$$^{-1}$, the $P_{[1,-1,0]}$ polarization curve first increase and then decrease within a narrow $k$ range (red ellipse). The inner upturn ($k\sim 0.48$ $\mathrm{\AA}$$^{-1}$) originates from a higher spin-up intensity, while the outer downturn ($k\sim 0.55$ $\mathrm{\AA}$$^{-1}$) originates from a higher spin-down intensity (inset of Fig. \ref{Fig2}\textbf{d}). Momentum splitting for the two rings is determined to be $\Delta k_{[110]} \sim 0.063$ $\mathrm{\AA}$$^{-1}$. Combined with their effective mass $m^*\sim4.8 m_e$, we estimated the Rashba coefficient to be $\alpha_\mathrm{R} = \hbar^2\Delta k_{[110]}/2m^* \sim 5 \times 10^{-12}$ eV m, which is about 5 times smaller than that in the Au(111) surface state \cite{LaShell}. Although the $\beta$ band spin splitting to the left of $\Gamma$ is not apparent in the raw polarization curve, standard analysis (Supplementary Information) reveal the in-plane spin direction for the inner contour, the results of which are shown as numbers in Fig. \ref{Fig2}\textbf{b}.

The unique spin fine structure resolved for the $\beta$ band agrees qualitatively with a theoretical model where finite Coulomb $U$ and surface termination give rise to Rashba-like in-plane spin texture (see Supplementary Information for details). In the case of $U = 0$, due to spin orbit coupling (SOC) and inversion symmetry breaking, both the $\alpha$ and $\beta$ bands develop their surface counterpart which exhibit similar bulk band dispersion but finite spin splitting and a small degree of spin polarization. The directions of the spin are strictly in plane ($S_z = 0$) and obey $E(k,\uparrow) = E(-k,\downarrow)$ due to time reversal symmetry. When $U$ is turned on, AFM order is obtained self-consistently with finite staggered magnetic moments along the $z$-direction. The time reversal symmetry is then broken and a finite $S_z$ component is obtained for each state (Supplementary Information). In our SR-ARPES measurements, we observed the in-plane Rashba-like spin texture and signature for non-zero $S_z$ component of the $\beta$ band (Supplementary Information). The presence of spin polarization reconfirms the surface-dominated signal of the ARPES data, since this splitting is not expected in the bulk electronic structure. All aspects of the data taken together, our experimental observation and first principles calculations suggest an interplay between SOC and the (bulk and/or surface) AFM order. Final state photoelectron spin effects reported in strong spin-orbit coupled topological insulators are minimized here, since the spin-orbit splitting and effective coupling in this system is even weaker than that in gold.

Such unique surface spin modulation is previously unobserved for a system with both strong SOC and onsite Coulumb interaction. These spin correlated surface electronic states are not observed in the high-$T_c$ superconducting cuprates, which indicates strong dissimilarity of the surface modified electronic states between the high-$T_c$ superconductors and the layered iridates, unlike what is predicted from theory \cite{Senthil}. Our finding adds an additional variable (or tunability) to the electronic phase diagram of the iridates. Since such surface related behavior are not observed in Sr$_2$IrO$_4$ or SrIrO$_3$, it is likely that the proximity to the bulk criticality contributes to the surface modification.

In Fig. \ref{Fig3} we further show $k_z$ dispersion analysis for Sr$_3$Ir$_2$O$_7$ with photon energies ranging from 25 eV to 80 eV. From the raw dispersive pattern and the associated momentum distribution curves (MDCs) (Figs. \ref{Fig3}\textbf{a}-\textbf{b}), we find that the resolved bands show little $k_z$ dispersion, as both the $\alpha$ and $\beta$ band form nearly vertical lines in the $k_{\|}$-$k_z$ plane along both $\Gamma$-$X$ and $\Gamma$-$M$ directions, which is evident for a quasi two dimensional electronic structure. On the other hand, our detailed analysis shown in Figs. \ref{Fig3}\textbf{c}-\textbf{d} reveals that the $\alpha$ band does show some extent of periodic $k_z$ dispersive pattern, which is required by the symmetry of the AF Brillouin zone. In Fig. \ref{Fig3}\textbf{d}, we show that the $Y$-$X$-$Y$ segment of the AF zone edge changes from a vertical line at $k_z = 0$, to a single dot $P$ at $k_z = \pi/c$, and finally to a horizontal line at $k_z = 2\pi/c$. As a result, any band close to this segment will have to change from a $k_{[1,-1,0]}$ elongated shape at $k_z = 0$ to a $k_{[110]}$ elongated shape at $k_z = 2\pi/c$. This is consistent with our ARPES results shown in Fig. \ref{Fig3}\textbf{c}, where the Fermi mappings done with $h\nu = 35$, 50 and 25 eV roughly correspond to the situation at $k_z=0$, $\pi/c$ and $2\pi/c$, respectively. To clarify our observations from Figs. \ref{Fig1} and \ref{Fig3}, a schematic constant energy map close to the top of the $\alpha$ band is presented in Fig. \ref{Fig3}\textbf{f} associated with the AF Brillouin zone (Fig. \ref{Fig3}\textbf{e}). From this figure, one can see that although some weak $k_z$ dispersion is discernable, the electronic structure of Sr$_3$Ir$_2$O$_7$ is mostly two dimensional. It is very important to note here that such finite $k_z$ dispersion is \textit{not excluded} for surface-related bands, especially for systems with small insulating gaps, since the surface bands can in principle penetrate deeper into the bulk and thus respect the symmetry of bulk electrons. The \textit{spin textured behavior} together with the observed weak $k_z$ dispersion in these states rules out the possibility that the $\alpha$ and $\beta$ bands are purely bulk bands.

From the data presented in Fig. \ref{Fig2}, we have experimentally investigated the presence of surface spin fine structure in Sr$_3$Ir$_2$O$_7$. The physical significance of the observed spin texture can be understood by comparing it to the well-known surface state found in topological insulators (TIs), as shown in Fig. \ref{Fig4}\textbf{a}. At the surface of a TI, a single gapless spin helical two dimensional electron gas (2DEG) presents as long as time reversal symmetry preserves in the system. Intrinsic interaction between the Dirac fermions and lattice phonons is found to be weak \cite{Batanouny}. At the surface of an AF-ordered correlated TMO, there exists a two dimensional electronic state (2DES) showing a unique spin fine structure. It is instructive to propose (see Fig. \ref{Fig4}\textbf{b}, adapted from Ref. \onlinecite{Moon_PRL}) that these spin polarized states reside only in the close vicinity of the MIT, where the bulk band gap is so small such that it is unstable at the crystal surface where lattice relaxation and band bending occur, which then give rise to an effective electric field along the $z$-direction. The spin helical surface state in TI and the spin-textured surfaces in TMOs near bulk criticality are examples of novel surface electronic structures that signify distinct phases of two dimensional condensed matter. In order to harness its exotic behavior, surface-to-bulk ratio can be increased by nano-structuring the sample in future layer-by-layer MBE growth techniques. In such thin films, anomalous quantum and Hall transport are expected in a strongly correlated setting.

In Fig. \ref{Fig4}\textbf{c}-\textbf{e} we present a first-principles GGA + $U$ band calculations that reveal a bulk electronic structure agreeing reasonably well with our measurements for high binding energies (detailed in Supplementary Information). In order to achieve such agreement, we set the Hubbard $U = 1.5$ eV and $\lambda_{\mathrm{SO}} = 1.7$ times the self consistent value in the calculation. Therefore our data places constraints on the magnitude of Coulumb $U$ and spin-orbit coupling experimentally realized in the material under study. The 12$^{\circ}$ rotation of the IrO$_6$ octahedra (Supplementary Information) gives rise to a Jahn-Teller type gap locating at 0.7 - 1.3 eV above $E_F$. The combined effect of $U$ and $\lambda_{\mathrm{SO}}$ causes the opening of a partial gap at $E_F$ close to $\Gamma$; this gap enlarges and becomes a $\sim180$ meV complete gap once AFM is added to the scenario, hence the term ``AFM gap''. One important difference between the theoretical bulk calculation and the experimentally observed electronic states lies around the $M$ point where the $\alpha$ and $\beta$ bands are observed to evolve up approaching $E_F$, making the circular electronlike shape of the $\beta$ band, while no bands are present within $E_b < 0.4$ eV in the bulk calculation (red arrow in Fig. \ref{Fig4}\textbf{c}). This critical difference enables the unique spin texture of the complete $\beta$ contours, indicating strong surface modification of the bands resolved by ARPES. Interestingly, our calculation optimized by fitting with experimental data on the real material also suggests that the real material possess a distinct, well-isolated surface state which energetically lies within the Jahn-Teller gap above the experimental Fermi level. This distinct surface state is robust even for a large variation within the parameter space in our calculation (Supplementary Information). Therefore, based on our experimentally optimized band calculation, we further predict the existence of an isolated correlated surface state in the highly $n$-doped Sr$_3$Ir$_2$O$_7$.

Although the layered nature of Sr$_3$Ir$_2$O$_7$ and the surface band dispersion (at higher binding energies) is similar to that of the bulk bands, surface effect can take place due to the residual interactions between two Ir$_2$O$_4$ blocks. The observed spin polarization is due to the surface effect since the bulk bands are spin degenerate. At the surface, the inversion symmetry is broken; the spin degeneracy is lifted due to the mutual effects of SOC and inversion symmetry breaking, resulting in one-to-one spin and momentum locking for the surface states in a band specific manner. Moreover, the time reversal symmetry is broken due possibly to AFM order so that $E(k,\uparrow)=E(-k,\downarrow)$ no longer holds. The spin splitting cannot be explained within the bulk band structure scenario since the bulk crystal structure of Sr$_3$Ir$_2$O$_7$ possess inversion symmetry, bulk bands must be spin degenerate even if they are quasi two dimensional. The surface effects are thus critically important to the interpretation of our data reflecting the near-surface ground state of iridate.

The surface electronic ground state of Sr$_3$Ir$_2$O$_7$, the $n = 2$ member of the Ruddlesden-Popper iridate series Sr$_{n+1}$Ir$_n$O$_{3n+1}$, is thus distinct from that in Sr$_2$IrO$_4$ which realizes a $J = 1/2$ Mott insulating state, and that of a topological insulator. Finite density of states is found at the Fermi level in Sr$_3$Ir$_2$O$_7$, with a gap-like spectral weight suppression on the order of 40 meV. Spin resolved ARPES data reveals a strong left-right imbalanced modulation on the surface, and finds a unique spin fine structure in one of the bands, which results from onsite Coulomb interaction and bulk and/or surface antiferromagnetism. These observations are evident for a strong interplay between spin-orbit coupling, bandwidth, long range magnetic order as well as surface formation in Sr$_3$Ir$_2$O$_7$. Our results provide unique insight for the quasiparticle interactions near the surface of this system, pointing toward the experimental realization of a novel two dimensional correlated electronic state, as well as paves the way for utilizing the observed surface electronic structure for future nano-structured quantum devices. The multiband nature and unique spin texture realized in this strong spin-orbit coupled material near the surface further suggests that the photon polarization-tuned emission of electron from the correlated surface can lead to harnessing the quantum entanglement of surface wavefunction for designing opto-spintronics \cite{Damascelli2} using the photon-based control of correlated electronic spin.

Work at Princeton and Princeton-led synchrotron-based measurements and the related theory at Northeastern University are supported by the Office of Basic Energy Sciences, US Department of Energy (grants DE-FG-02-05ER46200, AC03-76SF00098 and DE-FG02-07ER46352), and benefited from the allocation of supercomputer time at NERSC and Northeastern University's Advanced Scientific Computation Center. Work at Boston College is supported by NSF CAREER DMR-1056625. T.-R. C. is supported by the National Science Council and Academia Sinica, Taiwan and would like to thank NCHC, CINC-NTU, and NCTS, Taiwan for technical support. The Advanced Light Source is supported by the Director, Office of Science, Office of Basic Energy Sciences, of the U.S. Department of Energy under Contract No. DE-AC02-05CH11231. The Stanford Synchrotron Radiation Lightsource is supported by the U.S. Department of Energy under Contract No. DE-AC02-76SF00515. The Synchrotron Radiation Center is primarily funded by the University of Wisconsin-Madison with supplemental support from facility users and the University of Wisconsin-Milwaukee. We gratefully thank Sung-Kwan Mo, Jonathan D. Denlinger, Donghui Lu and Mark Bissen for instrumental support; Vidya Madhavan for fruitful discussion. C. L. acknowledges Peng Zhang, Takeshi Kondo, and Adam Kaminski for provision of data analysis software. M. Z. H. acknowledges Visiting Scientist support from LBNL and additional support from the A. P. Sloan Foundation.

\newpage

\begin{figure*}
\centering
\includegraphics[width=16cm]{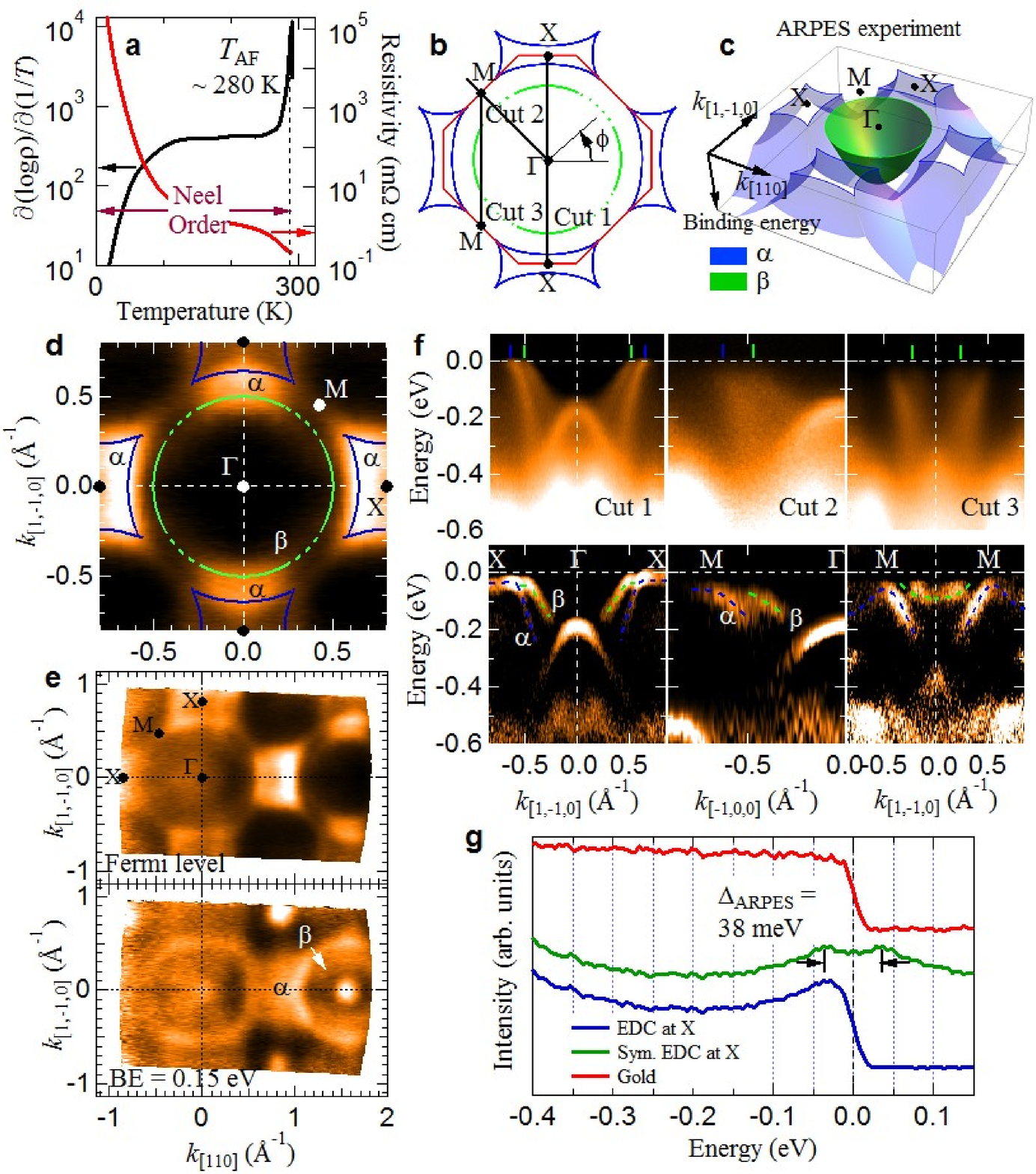}
\caption{\textbf{Surface electronic structure of Sr$_3$Ir$_2$O$_7$.} ARPES measurements are carried out at 14 K with 35 eV photons. \textbf{a}, Temperature dependence of the raw $a$-$b$ plane resistivity $\rho$ (right axis) and differentiated $a$-$b$ plane resistivity $\frac {\partial \mathrm{log} \rho}{\partial (1/T)}$ (left axis) for Sr$_3$Ir$_2$O$_7$. See Ref. \cite{Wilson_new} for details. \textbf{b}, Sketch of ARPES constant energy map close to $E_F$. Red lines mark the in-plane bulk Brillouin zone at $k_z = 0$. \textbf{c}, Sketch of the experiment-derived in-plane band dispersion at $k_z = 0$. \textbf{d}, ARPES Fermi mapping (see panel \textbf{e} for raw data). Data is two-fold symmetrized with respect to $k_{[110]}=0$. The $k_{[110]} \leq 0$ region represent the actual data. Blue and green colors represent the $\alpha$ and $\beta$ bands}
\end{figure*}
\addtocounter{figure}{-1}
\begin{figure*} [t!]
\caption{(Continued) respectively, and are used consistently over the paper. \textbf{e}, Raw ARPES mappings at 0 ($E_F$) and 0.15 eV binding energies. Note that data in the second Brillouin zone is used in panel \textbf{d} since the $\beta$ band is visible only in the second zone. \textbf{f}, Band dispersion along three $k$-$E$ cuts shown in \textbf{b}. Top: raw data; bottom: second derivative of raw data along the energy distribution curves (EDCs). \textbf{g}, Analysis of the gap value at $X$. The two peak structure of symmetrized EDC at $X$ near $E_F$ demonstrates the existence of a gap $\Delta_{\mathrm{ARPES}}(X) \sim 38$ meV.}\label{Fig1}
\end{figure*}

\clearpage

\begin{figure*}
\centering
\includegraphics[width=17cm]{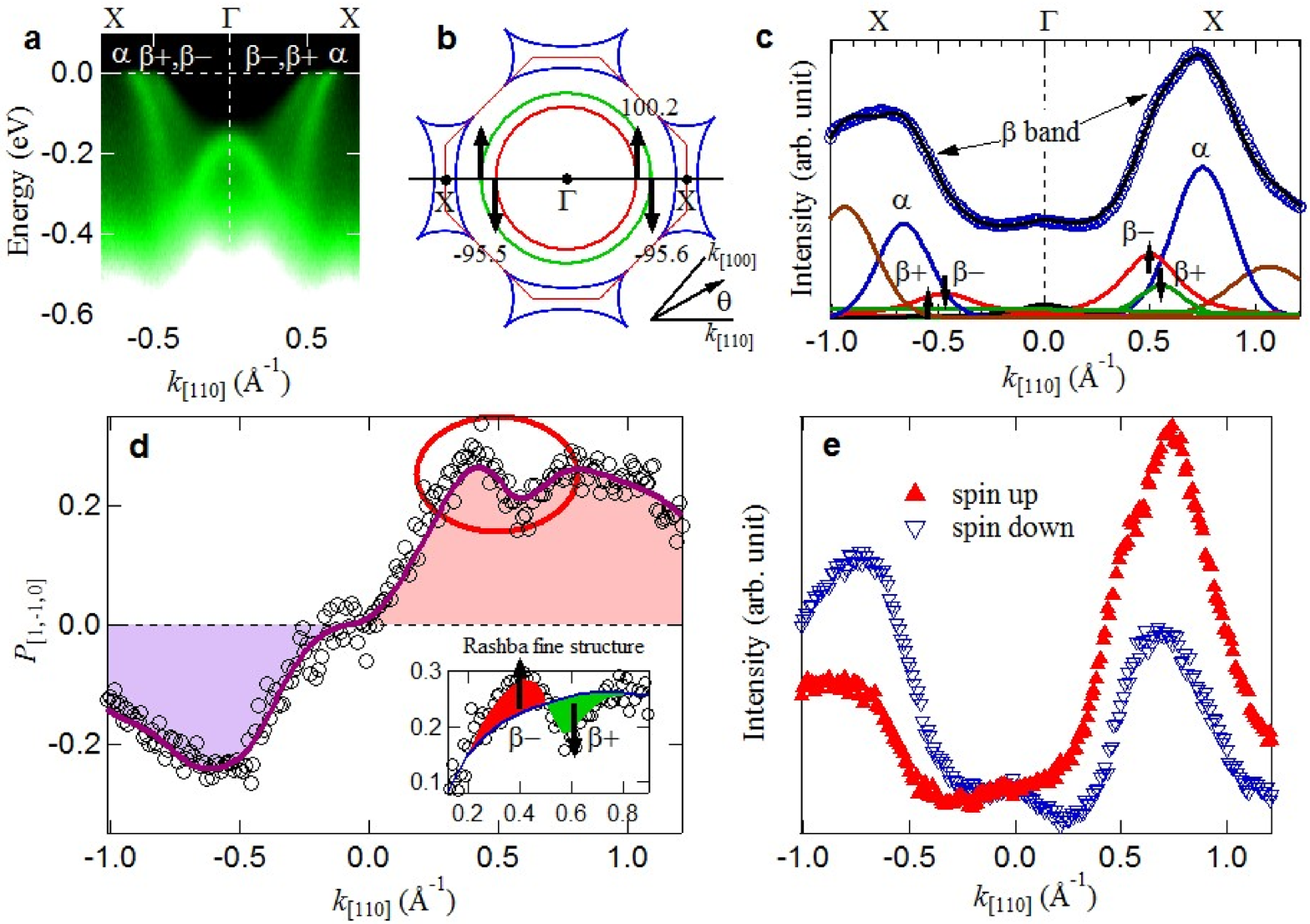}
\caption{\textbf{Spin polarization of electrons on the surface}. \textbf{a}, ARPES $k$-$E$ map along the $\Gamma$-$X$ direction. Data duplicated from Fig. \ref{Fig1}\textbf{f}. \textbf{b}, Summary of results from spin resolved ARPES measurements, together with notations of the AF Brillouin zone and cut direction in \textbf{c}-\textbf{e}. Blue contours represent the projected $\alpha$-band. Black arrows represent the Rashba-like in-plane spin polarization direction; numbers represent the experimental in-plane spin polarization angles ($\theta$). Some theoretically predicted spin components are not resolved in our measurements likely due to insufficient spin-polarized ARPES intensity. \textbf{c}, Spin integrated ARPES intensity along $\Gamma$-$X$ at a binding energy of $\sim50$ meV. Data is taken at $T = 20$ K with 35 eV photons. \textbf{d}, Polarization curve for spin component $S_{[1,-1,0]}$ along $\Gamma$-$X$. Insets show the Rashba-like in-plane spin fine structure of the $\beta$ band. \textbf{e}, Spin resolved ARPES intensity along the same direction as in \textbf{d}.
}\label{Fig2}
\end{figure*}

\begin{figure*}
\centering
\includegraphics[width=17cm]{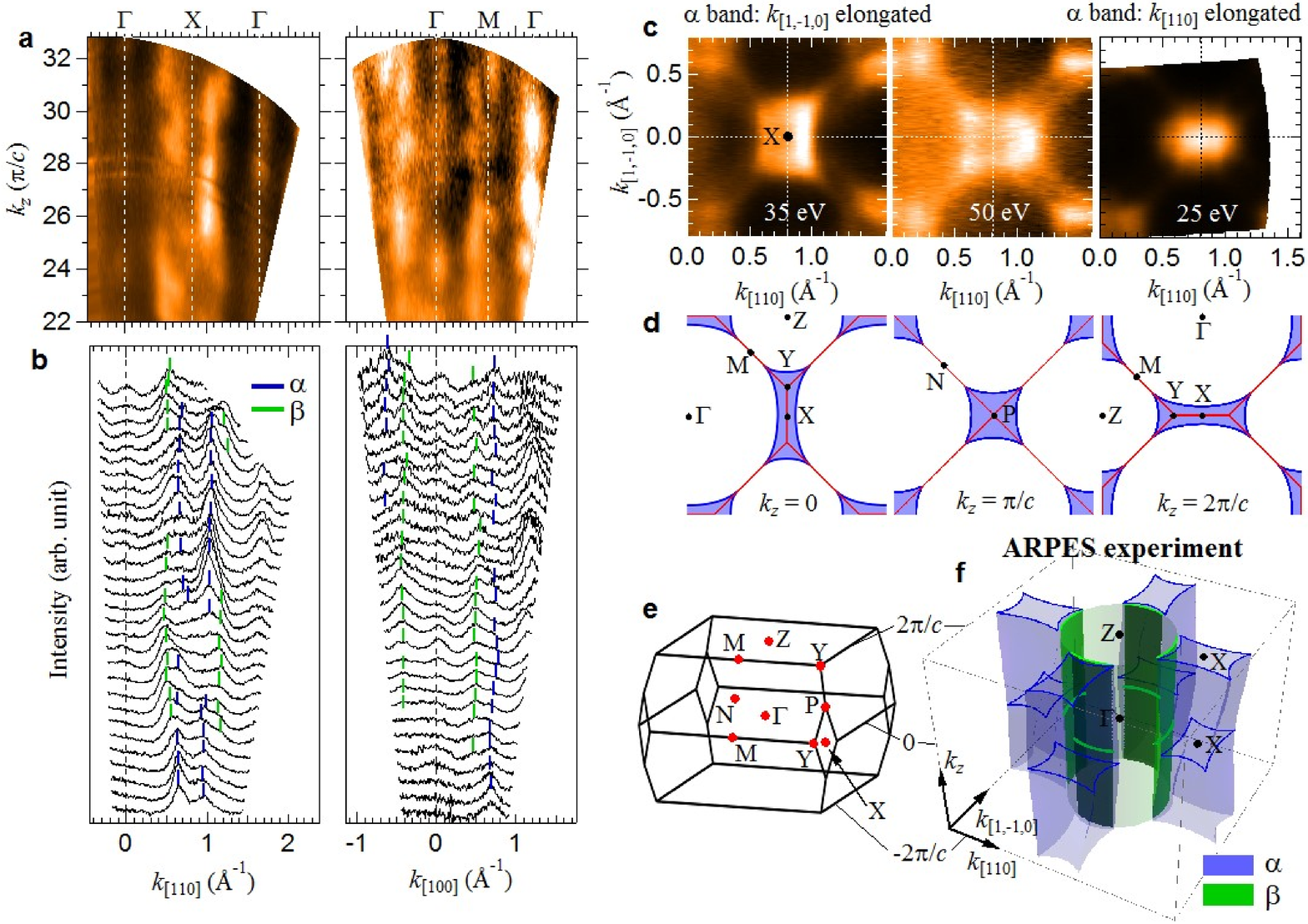}
\caption{\textbf{Quasi two dimensionality of the electronic states.} \textbf{a}, $k_z$ dispersion data ($40 < h\nu < 80$ eV) along two high symmetry directions. $k_z$ values are calculated using an inner potential $V_0=10$ eV. \textbf{b}, Momentum distribution curve (MDC) stacks of the corresponding panels in \textbf{a}. Blue and green bars mark the $\alpha$ and $\beta$ bands, respectively. \textbf{c}, ARPES Fermi mappings with 35, 50 and 25 eV photons, corresponding to $k_z$ values of $\sim 0$, $\sim \pi/c$ and $\sim 2\pi/c$, respectively. \textbf{d}, Sketches of experimental constant energy maps (CEMs) near $E_F$ for the corresponding photon energies in \textbf{c}. \textbf{e}, Brillouin zone and notation of high symmetry points. \textbf{f}, Schematics of the experiment-derived three dimensional CEM.}\label{Fig3}
\end{figure*}

\begin{figure*}
\centering
\includegraphics[width=15cm]{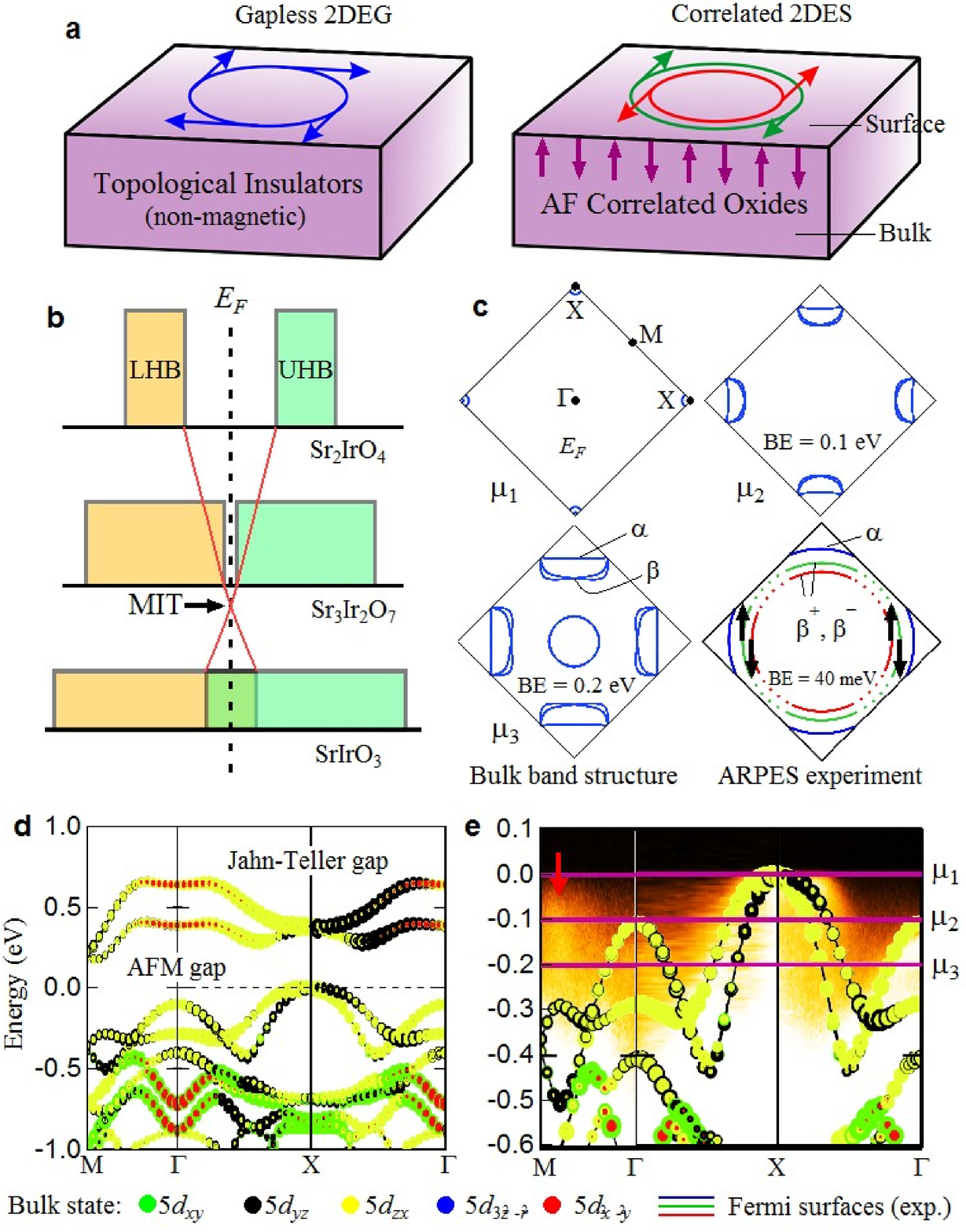}
\caption{\textbf{Layered iridate - GGA + $\textbf{U}$ theory vs. ARPES experiment}. \textbf{a}, Two types of two dimensional surface electronic structure - a single spin helical band found in topological insulators (left), and spin-orbit mediated Rashba-like spin splitting band found in our data on strongly correlated iridates (right). \textbf{b}, Schematic energy diagram of the Ruddlesden-Popper iridate series (adapted from Ref. \cite{Moon_PRL}). A metal-insulator transition (MIT) occurs when the upper Hubbard band (UHB) meets the lower Hubbard band (LHB) due to increased band width. \textbf{c}, Calculated constant}
\end{figure*}
\addtocounter{figure}{-1}
\begin{figure*} [t!]
\caption{(Continued) energy maps (CEMs) at different binding energies indicated by purple lines in \textbf{e}, versus the experimentally derived CEM at $E_B \sim 40$ meV (lower right panel). Black arrows show the Rashba-like fine texture of spin detected for the $\beta$ band. \textbf{d}, Results of band calculation (detailed in text). Solid circles with different colors and sizes represent the contribution of different Ir 5$d$ orbitals. \textbf{e}, Band calculation on top of ARPES band structure close to $E_F$ (same data set as Figs. \ref{Fig1} and \ref{Fig3}). Red arrow points to the $\beta$ band approaching $E_F$ near $M$ that is not present in the calculation.}\label{Fig4}
\end{figure*}
\end{document}